\documentstyle[12pt,thmsa,sw20aip]{article}

\input tcilatex

\begin{document}

\title{How to build a non-spreading wave packet in quantum mechanics}
\author{S. Bruce \\
Department of Physics\\
University of Concepcion\\
P.O. Box 4009\\
Concepcion, Chile}
\maketitle

\begin{abstract}
Quantum mechanics asserts that a wave packet must inevitably spread as time
progresses since the dispersion relation for the quantum waves is assumed to
be quadratic in the momentum $k.$ However, this assumption does not consider
the standard frequency Doppler shift formula of Galilean relativity. In this
article a non-dispersive wave packet is constructed by appropriately
considering the transformation rules between the laboratory and the
(particle's) rest inertial reference frames.\medskip 


theories.\bigskip

{\bf E-mail: sbruce@udec.cl}

{\bf Fax: (56-41) 22 4520.}
\end{abstract}

As is well known wave packets for particles of nonzero rest mass are
dispersed while they evolve in time. This is because wave-like properties
must be attributed to the quantum system, the wave vector ${\bf k}$ and the
frequency $\omega $ being related to the momentum and energy by 
\begin{equation}
{\bf p}\rightarrow \hbar {\bf k,\qquad }E\rightarrow \hbar \omega ({\bf k).}
\end{equation}
In the classical (non-relativistic) level the dependence of the energy on
the momentum must therefore be given by the familiar relationship $E({\bf %
p)=p}^{2}/2m_{0}$, where $m_{0}$ is the rest mass of the particle. The
corresponding dispersion relation for the {\it quantum waves} in the
free-particle case is therefore 
\begin{equation}
\omega ({\bf k)=}\frac{\hbar }{2m_{0}}{\bf k}^{2}.  \label{omega}
\end{equation}

Consider now a non-relativistic wave packet 
\begin{equation}
\Psi \left( {\bf x,}t\right) =\frac{1}{\left( 2\pi \hbar \right) ^{3/2}}\int
d^{3}k\exp \left[ i{\bf k\cdot x}\right] \Phi ({\bf k,}t).  \label{psi}
\end{equation}
We assume $\Psi $ is confined to a spatial region $V_{{\bf x}}$. We may
construct such a packet by means of a function $\Phi ({\bf k,}t)$ which is
concentrated in a volume $V_{{\bf k}}$ of $k$-space whose dimensions are
inversely proportional to $V_{{\bf x}}.$ Given the wave function at some
instant $t_{0}$ $(=0)$ throughout all the space, we can compute $\Psi $ at
any other future time. This follows from (\ref{psi}): 
\begin{equation}
\Psi \left( {\bf x,}t\right) =\frac{1}{\left( 2\pi \hbar \right) ^{3/2}}\int
d^{3}k\exp \left[ i\left( {\bf k\cdot x-}\omega ({\bf k})t\right) \right]
\Phi ({\bf k,}0).  \label{phi}
\end{equation}
Thus we can write 
\begin{equation}
\Psi \left( {\bf x,}t\right) =\int d^{3}x^{\prime }G({\bf x-x}^{\prime
},t)\Psi \left( {\bf x}^{\prime }{\bf ,}0\right) ,  \label{g}
\end{equation}
where 
\begin{equation}
G\left( {\bf x,}t\right) \equiv \frac{1}{\left( 2\pi \right) ^{3/2}}\int
d^{3}k\exp \left[ i({\bf k\cdot x-}\omega ({\bf k})t)\right] =\left( \frac{%
m_{0}}{2\pi i\hbar t}\right) ^{3/2}\exp \left[ \frac{im_{0}\mid {\bf x}\mid
^{2}}{2\hbar t}\right]  \label{gg}
\end{equation}
is the well known {\it free}-particle Green function\cite{got}. A particular
example of considerable interest is the one-dimensional gaussian wave
function 
\begin{equation}
\Psi \left( x{\bf ,}0\right) =A\exp (-(x-x_{0})^{2}/2a^{2})\exp
(ip_{x}x/\hbar ).
\end{equation}
From (\ref{g}) and (\ref{gg}) we get\cite{lei} 
\begin{eqnarray}
\Psi \left( x{\bf ,}t\right) &=&\pi ^{-1/4}\left( a+\frac{i\hbar t}{m_{0}a}%
\right) ^{-1/2}\exp \left[ \frac{ip_{x}x}{\hbar }-\frac{ip_{x}^{2}t}{%
2m_{0}\hbar }\right] \\
&&\times \exp \left[ -\frac{(x-x_{0}-p_{x}t/m_{0})^{2}(1-i\hbar /m_{0}a^{2})%
}{2(a^{2}+\hbar ^{2}t^{2}/m_{0}^{2}a^{2})}\right] .  \nonumber
\end{eqnarray}
The corresponding probability distribution is 
\begin{equation}
P(x,t)=\mid \Psi \left( x{\bf ,}t\right) \mid ^{2}=\pi ^{-1/2}\left( a^{2}+%
\frac{\hbar ^{2}t^{2}}{m_{0}^{2}a^{2}}\right) ^{-1/2}\exp \left[ -\frac{%
(x-x_{0}-p_{x}t/m_{0})^{2}}{(a^{2}+\hbar ^{2}t^{2}/m_{0}^{2}a^{2}}\right] .
\end{equation}

Notice that $G\left( {\bf x,}t\right) $ is the solution which evolves out of
the Dirac delta function $\delta ({\bf x})$ at $t=0.$ However, for $t>0$, $G$
will make $\Psi \left( {\bf x,}t\right) $ and $P({\bf x},t)$ to spread out
in space.

Let $K$ and $K^{\prime }$ be two inertial reference frames. The $K^{\prime }$
system has all its axes parallel to those of $K$ and moves with velocity $%
{\bf v}$ relative to $K.$ Assume $E,$ $p$ and $E^{\prime },$ $p^{\prime }$
to be the energy and momentum of a particle in the $K$ and $K^{\prime }$
frames, respectively. The Lorentz transformation for these quantities are 
\begin{equation}
E({\bf p)}={\bf v\cdot p+}\frac{E^{\prime }}{\gamma }\ ,  \label{e}
\end{equation}
\begin{equation}
{\bf p}={\bf p}^{\prime }{\bf +}\left( \gamma -1\right) \frac{{\bf p}%
^{\prime }\cdot {\bf v}}{v^{2}}{\bf v+}\gamma \frac{{\bf v}}{c^{2}}E^{\prime
}\ ,  \label{p}
\end{equation}
where $\gamma =\left( 1-v^{2}/c^{2}\right) ^{-1/2}.$ Let us now assume that $%
K^{\prime }$ is the rest reference frame of the particle. Equation (\ref{e})
becomes 
\begin{equation}
E({\bf p)}={\bf v\cdot p+}\frac{m_{0}c^{2}}{\gamma }\ .  \label{e2}
\end{equation}
It is interesting to note that this expression represents a correspondence
relation for the $E_{D}$-energy eigenvalues of the Dirac Hamiltonian. We can
see this by calculating the energy eigenvalues of the Dirac energy operator $%
H_{D}$ $=c{\bf \alpha \cdot p}+m_{0}c^{2}\beta $ in the energy-helicity
eigenstates $\mid E_{D},s>:$%
\begin{eqnarray}
E_{D} &\equiv &<E_{D},s\mid H_{D}\mid E_{D},s>  \label{e3} \\
&=&<E_{D},s\mid c{\bf \alpha \cdot p}+m_{0}c^{2}\beta \mid E_{D},s> 
\nonumber \\
&=&<E_{D},s\mid c{\bf \alpha \mid }E_{D}{\bf ,}s{\bf >\cdot p}+<E_{D},s\mid
\beta {\bf \mid }E_{D}{\bf ,}s{\bf >}m_{0}c^{2}\ .  \nonumber
\end{eqnarray}
Furthermore 
\begin{eqnarray}
&<&E_{D},s\mid c{\bf \alpha \mid }E_{D}{\bf ,}s{\bf >=}c^{2}\frac{{\bf p}}{%
E_{D}}={\bf v\ ,}  \label{mv} \\
&<&E_{D},s\mid \beta {\bf \mid }E_{D}{\bf ,}s{\bf >=}1{\bf /}\gamma \ , 
\nonumber
\end{eqnarray}
as can be easily checked \footnote{%
Here ${\bf \alpha }$ and $\beta $ are the usual Dirac matrices.}. Then from (%
\ref{e2}) and (\ref{e3}) we can make the correspondence $%
E\longleftrightarrow E_{D}.$ It is important to observe that a separation of
concepts is involved in (\ref{e2}) and (\ref{mv}). The term ${\bf v\cdot p}$
in (\ref{e2}) is the scalar product of the velocity ${\bf v}$ of the frame $%
K^{\prime }$, where the particle is at rest, and the momentum ${\bf p}$ of
the particle, both relative to the (say, laboratory) frame $K$. They are
velocities associated with different objects. Therefore, at the quantum
level, we have to distinguish between ${\bf p}/m_{0}$ and ${\bf v}$ although
they have the same {\it mean} values.

For the sake of simplicity, let us now assume the non-relativistic limit of (%
\ref{e2}). If $\mid {\bf v}/c\mid ^{2}\ll 1,$ we get 
\begin{equation}
E({\bf p})={\bf v\cdot p}-\frac{1}{2}m_{0}v^{2}+m_{0}c^{2}\ .
\end{equation}
Accordingly, the dispersion relation for the wave function (\ref{phi}) is 
\begin{equation}
\omega ({\bf k})={\bf v\cdot k}-\frac{\omega ^{\prime }}{2}\frac{v^{2}}{c^{2}%
}+\omega ^{\prime },\qquad \omega ^{\prime }=\frac{m_{0}c^{2}}{\hbar }\ .
\label{ome2}
\end{equation}
From (\ref{ome2}) and (\ref{phi}) we can build up a free-particle wave
packet that moves with group velocity ${\bf v=}\partial \omega ({\bf k}%
)/\partial {\bf k.}$ From (\ref{d}) and (\ref{ome2}) we find 
\begin{eqnarray}
\Psi \left( {\bf x,}t\right) &=&\frac{1}{\left( 2\pi \hbar \right) ^{3/2}}%
\int d^{3}k\exp \left[ i{\bf k\cdot (x-v}t)+i(\frac{\omega ^{\prime }}{2}%
\frac{v^{2}}{c^{2}}-\omega ^{\prime })t\right] \Phi ({\bf k,}0)  \label{wp}
\\
&=&\exp \left[ i\left( \frac{\omega ^{\prime }}{2}\frac{v^{2}}{c^{2}}-\omega
^{\prime }\right) t\right] \frac{1}{\left( 2\pi \hbar \right) ^{3/2}}\int
d^{3}k\exp \left[ i{\bf k\cdot (x-v}t)\right] \Phi ({\bf k,}0)  \nonumber \\
&=&\exp \left[ i\left[ \frac{\omega ^{\prime }}{2}\frac{v^{2}}{c^{2}}-\omega
^{\prime }\right] t\right] \Psi \left( {\bf x-v}t{\bf ,}0\right) \ . 
\nonumber
\end{eqnarray}
Therefore, we can write 
\begin{equation}
\Psi \left( {\bf x,}t\right) =\int d^{3}x^{\prime }{\cal G}({\bf x-x}%
^{\prime },t)\Psi \left( {\bf x}^{\prime }{\bf ,}0\right) \ ,  \label{ev}
\end{equation}
with 
\begin{equation}
{\cal G}({\bf x},t)=\exp \left[ -\frac{im_{0}c^{2}}{\hbar }\left( 1-\frac{%
v^{2}}{c^{2}}\right) t\right] \delta ^{(3)}({\bf x-v}t)\ ,
\end{equation}
the coresponding Green function. This shows that the wave packet (\ref{wp})
evolves without deformation with velocity ${\bf v}$ relative to the frame $K$%
.

Note that (\ref{ome2}) considers the Doppler effect of the quantum waves as
the result of the movement of the wave packet relative to an observer in $K.$
The consequence is the outcome of an unspreding wave packet.
Correspondingly, the probability density is a function that progresses
without deformation: 
\begin{equation}
P({\bf x},t)=\mid \Psi \left( {\bf x-v}t{\bf ,}0\right) \mid ^{2}\ .
\end{equation}

This work was supported by Direcci\'{o}n de Investigaci\'{o}n, Universidad
de Concepci\'{o}n, through grant P.I. 96.11.19-1.0. and by Fondecyt grant \#
1970995.

\end{document}